\documentstyle[proc,epsf]{article}

\def\Journal#1#2#3#4{{#1} {\bf #2}, #3 (#4)}

\def\PLB{{\em Phys. Lett.}  B}
\def\PRP{\em Phys. Rep.}

\def\ZPC{{\em Z. Phys.} C}

\def\Pom{{\bf I\!P}}              
\def\gsim{\mathrel{\rlap{\lower4pt\hbox{\hskip1pt$\sim$}}
    \raise1pt\hbox{$>$}}}         
\def\lsim{\mathrel{\rlap{\lower4pt\hbox{\hskip1pt$\sim$}}
    \raise1pt\hbox{$<$}}}         

\def\be{\begin{equation}}
\def\ee{\end{equation}}
\def\bea{\begin{eqnarray}}
\def\eea{\end{eqnarray}}

\begin{document}

\title{THE FORWARD CONE AND L/T SEPARATION IN DIFFRACTIVE DIS}

\author{A. V. PRONYAEV}

\address{Virginia Polytechnic Institute and State University,\\
Blacksburg, VA 24061-0435, USA} 


\maketitle\abstracts{LPS provides access to new fundamental observables:
the diffraction cone and azimuthal asymmetries.
Diffraction cone has a unique rise of $B_T$ from the exclusive limit to excitation of continuum $M^2 \approx Q^2$ which
is in striking contrast to experience with real photoproduction and hadronic diffraction.
Azimuthal asymmetry is large and pQCD calculable at large $\beta$
and can be measured with LPS. It allows testing of the pQCD prediction
of $L/T >> 1$}

\section{Helicity components of diffractive DIS}
The detection of leading protons $p'$ from diffractive DIS $ep\rightarrow
e'p'X$ gives access to several new observables: the diffraction slope
$B_{D}$ which quantifies the impact parameter properties of diffractive
DIS and new helicity structure functions. In this talk we report predictions~\cite{NPZ1}
for the nontrivial $\beta$-dependence of the diffraction slope and
suggest a new method~\cite{NPZ2} for measuring $R^{D}=d\sigma^{D}_{L}/d\sigma^{D}_{T}$
for diffractive DIS based on the azimuthal correlation of the $(e,e')$
and $(p,p')$ scattering planes.

The differential cross-section of the diffractive process $ep \rightarrow e'p'X$ reads
\bea
&&Q^{2}y{d\sigma(ep \rightarrow ep'X)\over dQ^2 dy dM^2 dp_{\perp}^2 d\phi}
= {1\over 2\pi}{\alpha_{em}\over  \pi}\left[\left(1-y+{y^2\over 2}\right){d\sigma^{D}_{T} \over dM^2 dp_{\perp}^2} 
+\left(1-y\right){d\sigma^{D}_{L} \over dM^2 dp_{\perp}^2}
\right.
\nonumber \\
&&\left.
+\left(1-y\right){d\sigma^{D}_{TT'} \over dM^2 dp_{\perp}^2}\cdot\cos{(2\phi)}
+\left(2-y\right)\sqrt{1-y}\;{d\sigma^{D}_{LT} \over dM^2 dp_{\perp}^2}\cdot\cos{(\phi)}\right],
\eea
where $p_{\perp}$ is the (p,p') momentum transfer, $\phi$ is the azimuthal angle between (e,e') and (p,p') scattering planes.

We focus on the $q\bar{q}$ excitation which dominates at large $\beta$.
Following the technique developed in~\cite{NZ}, we find (for the kinematical
variables see Fig.~1)
\be
{d\sigma_{i}^{D} \over dM^2 dp_{\perp}^2}=\frac{\alpha_{em}}{24\pi^2} \sum_{f} Z_f^2 \int d^2\vec{k}{1-J^2 \over 4J}\alpha_{S}^2
[h_i(z_{+})+h_i(z_{-})],
\ee
where $z_{\pm}={1\over 2}(1\pm J)$, $J=\sqrt{1-4{k^2+m_f^2\over M^2}}$,
$h_{T}=[1-2z(1-z)]\vec{\Phi}_1^2+m_f^2 \Phi_2^2$ ,
$h_{LT}\cdot\cos{(\phi)}=2z(1-z)(1-2z)Q(\vec{\Phi}_1\vec{t})\Phi_2$ ,
$h_{L}=4z^2(1-z)^2 Q^2 \Phi_2^2$ ,
$h_{TT'}\cdot\cos{(2\phi)}=2z(1-z)[\vec{\Phi}_1^2-2(\vec{\Phi}_1\vec{t})^2]$
and $m_{f}$ is the quark mass.

The helicity amplitudes $\vec{\Phi_{1}}$, $\Phi_{2}$ equal
\be
\Phi_j=\int d^2\vec{\kappa}\frac{f\left(x_{\Pom},\vec{\kappa},\vec{p}_{\perp}\right)}{\kappa^4}
[D_{j}(\vec{r}+\vec{\kappa})+D_{j}(\vec{r}-\vec{\kappa})
-D_{j}(\vec{r}+{\vec{p}_{\perp}\over 2})-D_{j}(\vec{r}-{\vec{p}_{\perp}\over 2})],
\ee
where $\vec{D}_{1}\left(\vec{r}\right)=\vec{r}\cdot D_{2}\left(\vec{r}\right)=\vec{r}/(\vec{r}^2+\epsilon^2)$, $\vec{r}=\vec{k}-({1\over 2}-z)\vec{p}_{\perp}$ and $f\left(x_{\Pom},\vec{\kappa},\vec{p}_{\perp}\right)$ is the off-forward unintegrated gluon
density.
\begin{figure}
\epsfxsize=4.5in
\epsffile{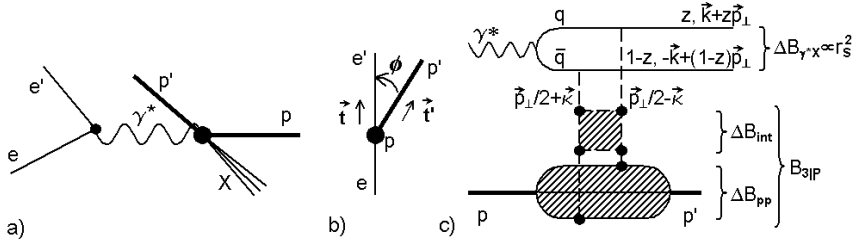}
\caption{a) Diffractive DIS, b) The definition of the azimuthal angle $\phi$, 
c) One of the pQCD diagrams for diffractive excitation of the $q\bar{q}$ state. 
\label{fig:diag}}
\end{figure}
Following~\cite{NZ,NNPZ,GNZ} we find
\be
\Phi_j \propto \int^{\bar{Q}^2} {d\kappa^{2} \over \kappa^2} f(x_{\Pom},\kappa^2,p_{\perp}^2)
= G(x_{\Pom},\bar{Q}^2,p_{\perp}^2)\approx G(x_{\Pom},\bar{Q}^2)\cdot (1-{1\over 2}B_{3\Pom} p_{\perp}^2).
\ee
where the pQCD hardness scale equals
\be
\bar{Q}^2=(k^2+m_f^2)\left(1+{Q^2\over M^2}\right)={k^2+m_f^2 \over 1-\beta}.
\ee
In Eq.~(4) we parameterize the small-$p_{\perp}^2$ dependence by the diffraction slope $B_{3\Pom}$
which comes from the proton vertex and gluon propagation effects (see Fig.~1c).
We emphasize that $B_{3\Pom}$ depends neither on $\beta$ nor flavor.

For excitation of heavy flavours we have the fully analytic
results (for the discussion of twist-4 $F_T$ and $F_{TT'}$ see~\cite{NPZ2,BGNPZ})
\bea
&&F_{T}^{D(4)}=
\frac{2\pi e_f^2}{9\sigma_{tot}^{pp}}
\frac{\beta(1-\beta)^2}{m_f^2}[
(1-B_{3\Pom}p_{\perp}^2)(3+4\beta+8\beta^2)
\nonumber \\
&&+\frac{p_{\perp}^2}{m_f^2}\frac{1}{10}(5-16\beta-7\beta^2-78\beta^3+126\beta^4)] \bar{G}_{T}^2
\nonumber \\
&&F_{L}^{D(4)}=
\frac{2\pi e_f^2}{9\sigma_{tot}^{pp}}
\frac{12\beta^3}{Q^2}[(1-B_{3\Pom}p_{\perp}^2) 2(1-2\beta)^2 \bar{G}_{L}^2
\nonumber \\
&&+\frac{p_{\perp}^2}{m_f^2}(1-\beta)(1-7\beta+23\beta^2-21\beta^3) \bar{G}_{T}^2 ]
\nonumber \\
&&F_{LT}^{D(4)}={p_{\perp}\over Q}\cdot
\frac{2\pi e_f^2}{9\sigma_{tot}^{pp}}
\frac{\beta^2 (1-\beta)}{m_f^2}[(1-B_{3\Pom}p_{\perp}^2)12\beta^2 (2-3\beta)
\nonumber \\
&&+\frac{p_{\perp}^2}{m_f^2}\frac{1}{20}(1-\beta)(2+7\beta+12\beta^2-483\beta^3+672\beta^4)] \bar{G}_{T}^2.
\eea
where $\bar{G}_{T,L}= \alpha_s(\bar{Q}^2_{T,L}) G(x_{\Pom},\bar{Q}^2_{T,L})$ and 
$\bar{Q}^2_T ={m_f^2\over 1-\beta}$, $\bar{Q}^2_L = {Q^2\over 4\beta}$.
The principal point is that $F_T$ and $F_{LT}$ are dominated by the aligned jet configurations,
$k^2 \sim m_f^2$, whereas $F_L$ comes from the large $k^2$ jets, $k^2 \sim {M^2\over 4}$.
In our calculations for light flavours we use the parameterization
for the soft glue from~\cite{BGNPZ}.

\section{Azimuthal asymmetry and $L/T$ separation}

In contrast to the inclusive DIS where $R=\sigma_L/\sigma_T \ll 1$, in
diffractive DIS  pQCD predicts~\cite{GNZ} $R^{D} \gg 1$ for $\beta > 0.9$,
despite the fact that $F_L$ is of higher twist.
Because neither the proton nor electron energy will be
changed at HERA, one must exploit the azimuthal asymmetry $A_{LT}^{D(4)}=F_{LT}^{D(4)}/(F_{T}^{D(4)}+F_{L}^{D(4)})$.
The key observation is that both $F_{LT}$ and $F_{T}$
come from the aligned jet configurations and the $LT/T$ ratio is model independent
\be
R^{D(4)}_{LT/T}=\frac{F_{LT}^{D(4)}}{F_{T}^{D(4)}}={p_{\perp}\over Q}\cdot\frac{12\beta^3\left(2-3\beta\right)}{\left(1-\beta\right)\left(3+4\beta+8\beta^2\right)}. 
\ee
Consequently, the measurement of $A_{LT}^{D(4)}$ amounts to the measurement of $R^{D(4)}\equiv F_{L}^{D(4)}/F_{T}^{D(4)}=R^{D(4)}_{LT/T}/A_{LT}^{D(4)}-1$.
The predicted asymmetry is quite substantial in the interesting region of $\beta \sim 0.9$
(Fig.~2), can be measured with the ZEUS and H1 leading proton spectrometers
and one can test the pQCD result $R^{D}\gg 1$ experimentally.
\begin{figure}
\hspace{1.4in}
\epsfxsize=2.3in
\epsffile{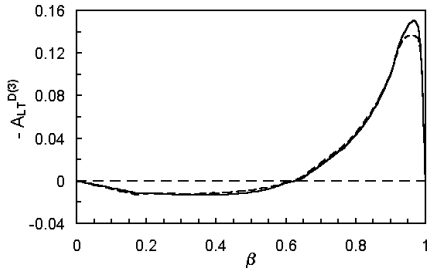}
\caption{Our prediction for the azimuthal asymmetry $A_{LT}^{D(4)}$ at 
$Q^2=100GeV^2$ and $x_{\Pom}=0.001$. The solid and dashed lines are for 
the GRV and MRS gluon structure functions, respectively.
\label{fig:asym}}
\end{figure}

\section{Peculiarities of the diffraction cone for diffractive DIS}

The diffraction slope $B_D$ is defined by the formula $d\sigma(ap\rightarrow XY) \propto \exp{(-B_D p_{\perp}^2)}$.
The experience with diffraction of hadrons and real photons can be summarized as follows.
One can write down an essentially model-independent decomposition $B_{D}=\Delta B_{aX} +\Delta B_{pY} + \Delta B_{int}$
in which $\Delta B_{int}$ interaction (exchange) range, and
$\Delta B_{aX}$ and $\Delta B_{pY}$ come from the (transverse) size of the
$aX$ and $pY$ transition vertices.
These contributions $\Delta B_{aX,pY}$ depend strongly on the excitation energy
in the $i\rightarrow j$ transition $\Delta M^{2} =m_{j}^{2}-m_{i}^{2}$
and vanish for excitation of the continuum~\cite{HAG}, $\Delta M^{2} \gsim 1\div 2 GeV^2$.

The experimental data on double diffraction dissociation $pp \rightarrow
XY$ into high mass states $X,Y$ give
$B_{D}\approx \Delta B_{int} \sim 1.5-2 GeV^{-2}$. In single diffraction
$ap\rightarrow Xp'$ into high mass states $B_{D} \approx \Delta B_{int}+\Delta B_{pp} \sim 6-7 GeV^{-2}$ is independent of the projectile $a=p, \pi, K, \gamma$.
It has been argued~\cite{NZ} that in the triple-pomeron regime of
diffractive DIS, $\beta \ll 1$, one must find $B_{D} \approx B_{3\Pom} \sim 6 GeV^{-2}$, which has indeed been confirmed by the ZEUS collaboration~\cite{ZEUS}.
Hereafter we focus on finite $\beta$, dominated by the $q\bar q$ excitation, $X=q\bar q$.

For diffractive DIS at finite $\beta$ the excited mass is large,
$M^2={1-\beta \over \beta}Q^2 \gg m_V^2$, hence the continuum is excited and naively
one would expect $\Delta B_{\gamma*X} \approx 0$, and $B_{D}\approx B_{3\Pom}$
independently of $\beta$.
Our principal finding is that this is not the case, $\Delta B_{\gamma*X}$
is large and varies substantially with $\beta$.

Our results for the small-$p_{\perp}^2$ of diffractive structure
functions are given by Eqs.(6). We focus on the transverse cross section
which dominates at $\beta < 0.9$. The component $\Delta B_{\gamma*X}$
comes from the term $\propto {p_{\perp}^2\over m_{f}^2}$.
These formulas are directly applicable for heavy flavour excitation.
In the diffraction excitation of light flavours there is a
sensitivity to the gluon structure function in the soft region, and
the rate of variation of the gluon structure function in the soft
region emerges as a scale instead of ${1\over m_{f}^2}$.  However, the
qualitative features of the $\beta$ dependence do not change from
heavy to light flavours. The numerical results are shown in Fig.~3.

The most striking prediction is the rise of $B_D$ when $\beta$
decreases from $\beta \sim 1$ to $\beta \sim {1\over 2}$. This rise
can be related to the rise of the scanning radius discussed in~\cite{GNZ}:
$r_S^2 \sim {1\over \bar{Q}^2_T} \sim {1-\beta \over m_f^2}$. Numerically, at $\beta \sim 1/2$ we have $\Delta B_{\gamma*X} \sim {1\over 10m_f^2}$.
In excitation of small masses,  $4m_f^2 \ll M^2 \ll Q^2$, i.e., $\beta \rightarrow 1$, we predict
a substantial drop of $B_{D}$, because here $\Delta B_{\gamma*X} \sim -{1\over 5m_f^2}$.
This is a legitimate pQCD domain because $\bar{Q}^2_T={Q^2\over 4}$ for $M^2 \sim 4m_f^2$. 
Very close to the threshold
\be
F_{T}^{D(4)}(p_{\perp}^2,x_{\Pom},v,Q^{2})=\frac{128\pi e_f^2}{3\sigma_{tot}^{pp}}
{m_f^2 \over Q^4} v [(1-B_{3\Pom}p_{\perp}^2)+
{p_{\perp}^2 \over 6m_f^2}v^2] \bar{G}_{T}^2,
\ee
where $v=\sqrt{1-{4m_f^2\over M^2}}$. In the spirit of exclusive-inclusive duality, $B_D=B_{3\Pom}$
can be related to the diffraction slope for V(1S) vector meson production,
whereas the drop of $B_D$ for somewhat higher masses correlates nicely with
the prediction~\cite{NNPZ} $B_{V'(2S)} \ll B_{V(1S)}$.
The experimental observation of this large-$\beta$ drop of $B_D$ is not easy
because of the masking effect of the longitudinal cross section for which
$B_D \approx B_{3\Pom}$.
\begin{figure}
\hspace{0.3in}
\epsfxsize=4.0in
\epsffile{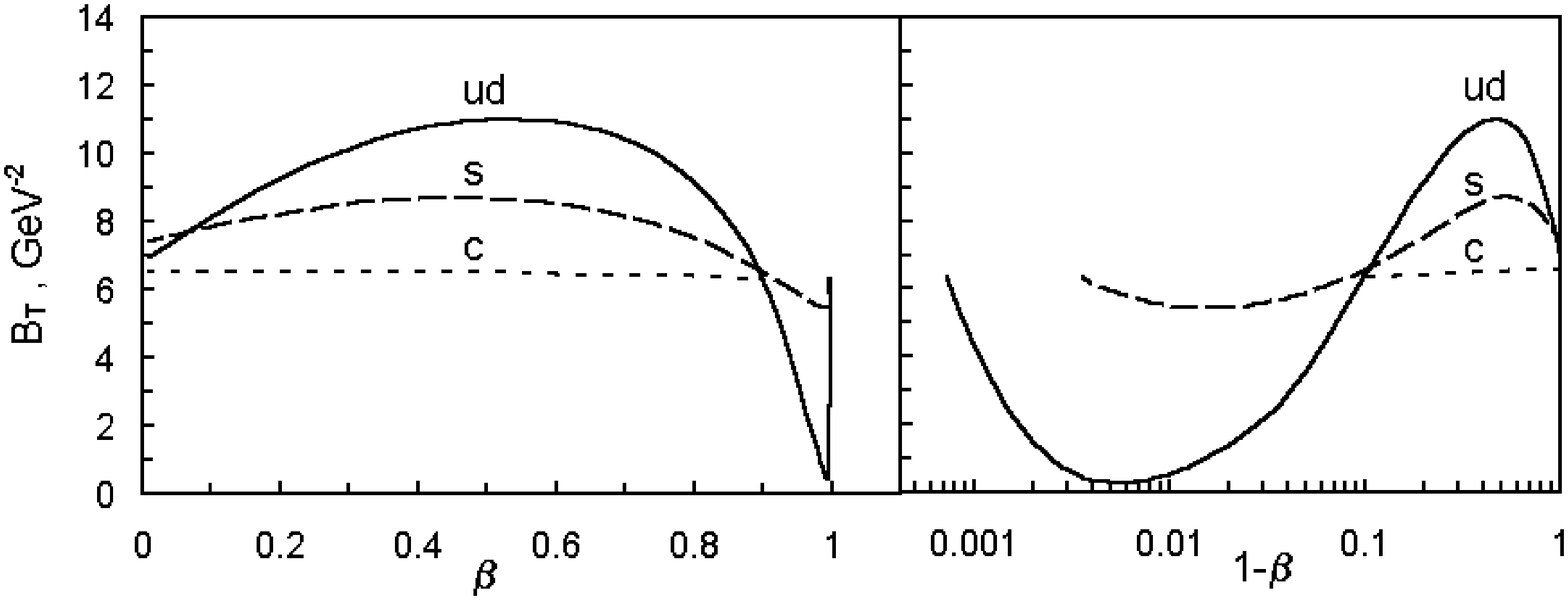}
\caption{Our predictions for the diffraction slope $B_{D}$ for the transverse
structure function at $Q^2=100GeV^2$ and $x_{\Pom}=0.001$.
\label{fig:slope}}
\end{figure}

\end{document}